\documentclass[preprint,showkeys,showpacs,preprintnumbers,amsmath,amssymb]{revtex4}
\usepackage{graphicx}
\usepackage{dcolumn}	% Align table columns on decimal point
\usepackage{bm}	       % bold math

\usepackage{amssymb, amsmath}
\usepackage{mathrsfs}  % required for $\lap$ := $\mathscr L$ ...
\usepackage{color}

\newcommand{\D}{\partial}
 \newcommand{\ve}[1]{{\bf #1}}
 \newcommand{\vve}[1]{{\bf { #1}}}
 \newcommand{\vveg}[1]{{\boldsymbol {#1}}}	% Greek matrix

 \newcommand{\const} {{\rm const}}

 \newcommand{\delt} {{\Delta t}}

\newcommand{\mat}[4]{ \bracket{\begin{array}{ll}
				#1  &#2\\
				#3  &#4
				\end{array}}}
\newcommand{\vect}[2]{ \bracket{\begin{array}{l}
				#1\\
				#2 
				\end{array}}}

\newcommand{\dt}[1]{\frac {d#1} {d t}}
\newcommand{\abs}[1]{\left|#1\right|}
\newcommand{\bracket}[1]{\left[#1\right]}

\DeclareMathSymbol{\R}{\mathbin}{AMSb}{"52}

 \newcommand{\DI}[1]{\frac {\D #1} {\D x_1}}
 
 \newcommand{\DIDI}[1]{\frac {\D^2 #1} {\D x_1^2}}

\begin{document}

   \title{\bf 
	A note on causation versus correlation}

        \author{X. San Liang}
	% \footnote {URL: http://people.deas.harvard.edu/$\sim$san/ }}

	\email{x.san.liang@gmail.com}
	\affiliation{Nanjing Institute of Meteorology, Nanjing, China}

	\author{Xiuqun Yang}
	\affiliation{School of Atmospheric Sciences,
	Nanjing University, Nanjing, China}

         \date{\today}

\begin{abstract}
{
   Recently, it has been shown that the causality and information flow
between two time series can be inferred in a rigorous and quantitative
sense, and, besides, the resulting causality can be normalized. 
A corollary that follows is, in the linear limit, causation implies
correlation, while correlation does not imply causation. Now suppose there
is an event $A$ taking a harmonic form (sine/cosine), and it generates through
some process another event $B$ so that $B$ always lags $A$ by a phase of 
$\pi/2$. Here the causality is obviously seen, while by computation the
correlation is, however, zero. This seemingly contradiction is rooted in
the fact that a harmonic system always leaves a single point on the
Poincar\'e section; it does not add information. That is to say, though
the absolute information flow from $A$ to $B$ is zero, i.e., $T_{A\to B}=0$, 
the total information increase of $B$ is also zero,
so the normalized $T_{A\to B}$, denoted as $\tau_{A\to B}$, 
takes the form of $\frac 00$.
By slightly perturbating the system with some noise,
solving a stochastic differential equation, 
and letting the perturbation go to zero,
it can be shown that $\tau_{A\to B}$ approaches 100\%,
just as one would have expected. 
}
\end{abstract}



\maketitle

%\section{A review}

Causal inference is a fundamental problem in scientific research.
	%lying at the heart of science. 
Recently it has been shown that the problem can be recast into the 
framework of information flow, another fundamental notion in
general physics which has wide applications in different disciplines 
(see \cite{Liang2016}), and hence put on a rigorous footing. The causality
between two time series can then be analyzed in a quantitative sense, and,
besides, the resulting formula is very concise in form. In the linear 
limit, it involves only the usual statistics namely sample
covariances\cite{Liang2014},
making the important and otherwise difficult problem an easy task.

%We begin by stating a principle or an observational fact about
%causality:
%	\begin{itemize}
%	\item[$\ $]
%   	{\it If the evolution of an event, say, $X_1$, is independent 
%	of another one, $X_2$, then the causality from $X_2$ to 
%	$X_1$ is zero.}
%	\end{itemize}
%Since it is the only quantitatively stated fact about causality, 
%all previous empirical/half-empirical causality formalisms have attempted 
%to verify it in applications. Considering its importance, it has been 
%referred to as the {\bf principle of nil causality} (Liang, 2016). 
%Recently, Smirnov (2013) systematically examined 
%the traditional formalisms, i.e., transfer entropy analysis and/or 
%Granger causality testing, 
%and found that they cannot verify the principle in a wide range of 
%situations; similar conclusion has been drawn by Lizier and Prokopenko (2010).
%We will see soon below that, within our framework, 
%this principle turns out to be a proven theorem.

To briefly review the theory, 
consider a two-dimensional continuous-time stochastic system
for state variables $\ve x = (x_1, x_2)$
	\begin{eqnarray}	\label{eq:stoch_gov}
	\dt {\ve x} = \ve F(\ve x, t) + \vve B(\ve x, t) \dot {\ve w},
	\end{eqnarray}
where $\ve F = (F_1, F_2)$ 
may be arbitrary nonlinear functions of $\ve x$ and $t$,
$\dot {\ve w}$ is a vector of white noise, and $\vve B  = (b_{ij})$ 
is the matrix of perturbation amplitudes 
which may also be any functions of $\ve x$ and $t$.
Here we adopt the convention in physics and do not distinguish
deterministic and random variables; in probability theory, they are
ususally distinguished with capital and lower-case symbols.
Assume that $\ve F$ and $\vve B$ are both differentiable with respect
to $\ve x$ and $t$. 
Then the information flow from $x_2$ to $x_1$ (in nats per unit time) 
can be explicitly found in a closed form:
%
%	\begin{thm} {\rm (Liang, 2016)}\\
%	For the system (\ref{eq:stoch_gov}),
%	the rate of information flowing from $x_2$ to $x_1$ (in nats) is
	\begin{eqnarray}	\label{eq:T21}
	T_{2\to1} = -E \bracket{\frac1{\rho_1} \DI{(F_1\rho_1)}}
	    + \frac12 E \bracket{\frac1{\rho_1} 
		\DIDI {g_{11}\rho_1}},
	\end{eqnarray}
	where $E$ stands for mathematical expectation, and
	$g_{ii} = \sum_{k=1}^n b_{ik} b_{ik}$, 
	$\rho_i = \rho_i(x_i)$ is the marginal probability density function
	(pdf) of $x_i$. The rate of information flowing from $X_1$ to $x_2$
	can be obtained by switching the indices.
%	\end{thm}
If $T_{j\to i} = 0$, then $x_j$ is not causal to $x_i$; otherwise 
it is causal, and the absolute value measures the magnitude of the 
causality from $x_j$ to $x_i$.
For discrete-time mappings, the information flow is in much more
complicated a form; see \cite{Liang2016}.
 
%There is a nice property for the above information flow: % (Liang, 2008):
%	\begin{thm} {\rm (Principle of nil causality) (Liang, 2008)}\\
%	If in (\ref{eq:stoch_gov})
%	neither $F_1$ nor $g_{11}$ depends on $x_2$, 
%	then $T_{2\to1} = 0$.
%	\end{thm}
%Note this is the very {\it principle of nil causality}. 
%While with classical ansatz-like formalisms people strive to verify it
%(e.g., Smirnov, 2013), here, remarkably, it appears as a proven theorem.

In the case with only two time series (no dynamical system is given)
$X_1$ and $X_2$, 
%we have the following result (Liang, 2014):
% the causality between them can be estimated using 
% maximum likelihood estimation (Liang, 2014).
%	\begin{thm} \label{thm:L14} % {\rm (Liang, 2014)} \\
%	Given two time series $X_1$ and $X_2$, 
	under the assumption of a linear model with additive noise,
	the maximum likelihood estimator (MLE) of the rate of information 
	flowing from $X_2$ to $X_1$ is\cite{Liang2014}
	\begin{eqnarray}	\label{eq:T21_est}
	\hat T_{2\to1} = \frac {C_{11} C_{12} C_{2,d1} - C_{12}^2 C_{1,d1}}
			  {C_{11}^2 C_{22} - C_{11} C_{12}^2},
	\end{eqnarray}
	where $C_{ij}$ is the sample covariance
	between $X_i$ and $X_j$, 
	and $C_{i,dj}$ the sample covariance between $X_i$ and 
	a series derived from $X_j$ using the Euler forward differencing
	scheme:
	$\dot X_{j,n} = (X_{j,n+k} - X_{j,n}) / (k\delt)$, with $k\ge1$ 
	some integer.
%	\end{thm}
Note that Eq.~(\ref{eq:T21_est}) is rather concise in form; 
it only involves the common statistics, i.e., sample covariances. 
In other words, a combination of some sample convariances will give 
a quantiative measure of the causality
between the time series. This makes causality analysis, which otherwise 
would be complicated with the classical empirical/half-empirical methods, 
very easy. Nonetheless, note that Eq.~(\ref{eq:T21_est}) cannot replace
(\ref{eq:stoch_gov}); it is just the mle of the latter. Statistical
significance test must be performed before a causal inference is made based
on the computed $T_{2\to1}$. For details, refer to \cite{Liang2014}.

% endeavor

Considering the long-standing debate ever since Berkeley
(1710)\cite{Berkeley1710}
over correlation versus causation, we may rewrite (\ref{eq:T21_est})
in terms of linear correlation coefficients, which immediately
implies\cite{Liang2014}:
%	\begin{eqnarray}
%	T_{2\to1} = \frac r {1-r^2} 
%		    \parenth{r'_{2,d1} - rr'_{1,d1}},
%	\end{eqnarray}
% with 
%	$r = \frac {C_{12}} {\sqrt{C_{11} C_{22}}}$
% the correlation coefficient, and
%	$r'_{i,dj} = \frac {C_{i,dj}}
%		     {\sqrt{C_{ii} C_{jj}}}.$
% Observe that, if $r=0$, then $T_{2\to1} = 0$; but if $T_{2\to1}=0$,
% $r$ does not necessarily vanish. Contrapositively, this means that
% A direct corollary is that, in the linear sense\cite{Liang14},
	\begin{itemize}
	\item[\ ] {\it Causation implies correlation, but correlation does not
		  imply causation.}
	\end{itemize}

%Causality can be normalized in order to reveal the relative importance 
% of a causal relation. But the normalization is by no means
%as trivial as that for covariance, considering that
%information flow is asymmetric in direction
%($T_{2\to1} \ne T_{1\to2}$ in general), and, besides, there is
%no such a property like Schartz inequality which makes it possible for
%covariance to be normalized. In Liang (2015), a way of normalization is 
%given, which in the following we will revisit.
%
%
The above formalism has been validated with many benchmark systems
(e.g., \cite{Liang2016}) such as baker transformation, H\'enon\ map, 
Kaplan-Yorke map, R\"ossler system, etc.
% Particularly, Eq.~(\ref{eq:T21_est}) has been validated with touchstone 
% problems where the traditional Granger causality test and transfer 
% entropy analysis fail. 
% An example is the highly chaotic anticipatory system problem described
% in Hahs and Pethel (2011), which with (\ref{eq:T21_est}) turns out not 
% to be a problem at all.
It also has been successfully applied to the studies of
many real world problems such those in financial economics
(e.g., the ``Seven Dwarfs vs. a Giant'' problem\cite{Liang2015}), 
earth system science (e.g., the Antarctica mass balance
problem\cite{Vannitsem2019} and the global warming problem\cite{Stips2016}), 
neuroscience (e.g., the concussion
problem\cite{Hristopulos2019}), to name but a few.

Now suppose we have an dynamic event $A$ which drives another $B$. The 
former has a harmonic form, leading the latter 
by a phase of $90^o$.
That is to say, 
the time series resulting from the two are in quadrature.
Then the correlation between the two are zero. However, since $A$ causes
$B$, the result is apparently in contradiction to the corollary that 
"causation implies correlation."

% The word quadrature has three incompatible meanings. Integration by
% quadrature either means solving an integral analytically, or solving of
% an integral numerically (e.g., Gaussian quadrature, Newton-Cotes
% formulas). Ueberhuber (1997, p.71) uses the word "quadrature" to mean 
% numerical computation of a univariate integral, 
% and "cubature" to mean numerical computation of a multiple integral.
%
% The word quadrature is also used to mean "squaring": the construction of
% a square using only copass and straightedge which has the same area as a
% given geometric figure. If quadrature is possible for a plane figure, it
% is said to be quadrable.
%
% Quadrature phase. Ocillations that are said to be in quadrature if they
% are separated in phase by 90 deg (pi/2).

%\section{The solution}

The problem can be more formally stated with the harmonic system:
	\begin{eqnarray}	\label{eq:harmonic}
	\dt {\ve x} = \ve F(\ve x, t) = \vve A \ve x 
	= \mat {a_{11}} {a_{12}} {a_{21}} {a_{22}} 
	  \vect {x_1} {x_2} 
	= \mat 0 {-1} 1 0 \vect {x_1} {x_2}.
	\end{eqnarray}
If the system is initialized with $x_1(0) = 1$, $\dot x_1(0) = 1)$, 
the solution is, $x_1 = \cos t$, $x_2 = \sin t$. 
So the population covariance $\sigma_{12} = \int \cos t \sin t dt = 0$
(the integral is taken over one or many periods). This yields an
information flow from $x_2$ to $x_1$:
	\begin{eqnarray}
	T_{2\to1} = a_{12} \frac {\sigma_{12}} {\sigma_{11}} =  0.
	\end{eqnarray}

Fundamentally the above problem arises from the fact that it is a
deterministic system. In Granger causality test, this case has been
explicitly excluded, as in such case the trajectories do not form appropriate
ensembles in the sample space.
For a harmonic series, on a Poincare section it is only one single point;
so the total information does not accure. 
% Indeed, in the sample space, the above sine/cosine
% solution forms only one sample path; probability collaps at this 
% singular point. The marginal entropy of $x_1$ is hence zero, and it 
% does not change as time moves on. 
If the total information does not change, the information 
flow to $x_1$ must also vanish. 
However, the vanishing information flow
does not mean that there is no influence of $x_2$ on $x_1$. 
As we argued in Liang (2015), the so-obtained information must be normalized,
just as covariance needs to be normalized into correlation, for one to
assess the causal influence.
Here if the normalizer is zero, the problem becomes something like
$\frac 00$ in calculus. We may then approach it by taking the
limit. Specifically, we may approach it by enlarging the sample space 
slightly, i.e., by adding some stochasticity to the system, then take 
the limit by letting the stochastic perturbation amplitude go zero.

% As we stated in Liang (2015), the absolute value of an information flow
% actually does not tell how large the causality is. The importance of a
% variable to another can only be measured by its relative value. This
% motivates us to normalize the information flow. 

By Liang (2015), the normalizer for $T_{2\to1}$ is
	\begin{eqnarray}
	Z_{2\to1} = \abs{T_{2\to1}} + \abs{\dt {H_1^*}} 
		  + \abs{\dt {H_1^{\rm noise}}},
	\end{eqnarray}
where on the right hand side, the second term is the contribution from
$x_1$ itself, and the third term the contribution from noise.
In Liang (2015), it has shown that $\dt {H_1^*}$ is a Lyapunov
exponent-like, phase-space stretching rate, and $\dt {H_1^{noise}}$
a noise-to-signal ratio. In this problem, we do not have noise taken into
account. But in reality, noise is ubiquitous. We may hence view a
deterministic system as a limit or extreme case as the amplitude of
stochastic perturbation goes to zero. For this case, we add to
(\ref{eq:harmonic}) a stochastic term:
	\begin{eqnarray}
	\dt {\ve x} = \vve A \ve x + \vve B \dot{\ve w},
	\end{eqnarray}
where ${\ve w}$ is a vector of standard Wiener processes. 
For simplicity, let the perturbation amplitude $\vve B$ a constant matrix.
Further let $\vve G = \vve B \vve B^T$, with elements 
	$$(g_{ij}) = \sum_{k=1}^2 b_{ik} b_{jk}.$$
Liang (2008) established that
	\begin{eqnarray}
	&& \dt {H_1^*} = a_{11} = 0,	\\
	&& \dt {H_1^{\rm noise}} = \frac12 \frac{g_{11}} {\sigma_{11}}.
	\end{eqnarray}
So in this case, the normalized flow from $x_2$ to $x_1$ is 
	\begin{eqnarray}
	\tau_{2\to1} = \frac {a_{12} \frac{\sigma_{12}} {\sigma_{11}} }
	                {\abs{a_{12} \frac{\sigma_{12}} {\sigma_{11}} }
			+ 0
			+ \abs{\frac 12 \frac {g_{11}} {\sigma_{11}}} }
	= \frac {-\sigma_{12} }
		{\abs{\sigma_{12}} + \frac {g_{11}} 2}.
	\end{eqnarray}

Now for stochastic equation, 
the covariance matrix $\vveg\Sigma$ evolves as
	\begin{eqnarray}
	\dt {\vveg\Sigma} = \vve A \vveg\Sigma + \vveg\Sigma \vve A^T 
			+ \vve B\vve B^T
	                = \vve A \vveg\Sigma + \vveg\Sigma \vve A^T 
			+ \vve G
	\end{eqnarray}
Expanding, this is
	\begin{eqnarray*}
	\dt\ \mat {\sigma_{11}} {\sigma_{12}} {\sigma_{12}} {\sigma_{22}}
	= \mat {-\sigma_{12}} {-\sigma_{22}} {\sigma_{11}} {\sigma_{12}}
	+ \mat {-\sigma_{12}} {\sigma_{11}} {-\sigma_{22}} {\sigma_{12}}
	+ \mat {g_{11}} {g_{12}} {g_{12}} {g_{22}}.
	\end{eqnarray*}
We hence obtain the following equation set:
	\begin{eqnarray*}
	&& \dt {\sigma_{11}} = -2\sigma_{12} + g_{11},	\\
	&& \dt {\sigma_{12}} = -\sigma_{22} + \sigma_{11} + g_{12},\\
	&& \dt {\sigma_{22}} = 2\sigma_{12} + g_{22}.
	\end{eqnarray*}
Solving, we get
	\begin{eqnarray*}
	\frac {d^2 \sigma_{12}} {dt^2} = -4\sigma_{12} 
		+ (g_{11}-g_{22}+g_{12}).
	\end{eqnarray*}
So the solution is
	\begin{eqnarray*}
	\sigma_{12} = C_1 \cos 2t + C_2 \sin2t 
		    + \frac12 (g_{11} - g_{22} + g_{12}) t^2.
	\end{eqnarray*}
If $\sigma_{12}(0) = 0$, $\dot\sigma_{12}(0) = 0$, then the integration
constants $C_1 = C_2 = 0$. So
		\begin{eqnarray*}
		\tau_{2\to1} = \frac {-\sigma_{12}}
			       {\abs{\sigma_{12}} + \frac12 g_{11}}
		= \frac {-1} {1 + \frac {g_{11}} {(g_{11}-g_{22}+g_{12})t^2} }
		\end{eqnarray*}
Two cases are distinguished:
	\begin{itemize}
	\item[Case I] $g_{12} - g_{22} = \const \ne 0$. 
			$$\lim_{g_{11}\to0} \tau_{2\to1} \to -1.$$
	\item[Case II] $g_{12} - g_{22} = 0$.
		$$\lim_{g_{11}\to0} \tau_{2\to1} \to \frac {-1} {1 + 1/t^2}.$$
		As $t$ goes to infinity, $\tau_{2\to1}$ also approaches -1.
	\end{itemize}
If initially there exists some covariance, say, $\sigma_{12}(0) = c$, then
$C_1 = c$, and hence
		\begin{eqnarray*}
		\tau_{2\to1} = \frac {-1} 
		{1 + \frac {g_{11}} {c\cos 2t + (g_{11}-g_{22}+g_{12})t^2} }.
		\end{eqnarray*}
In this case, as $g_{11}\to 0$, we always have $\tau_{2\to1} \to -1$.
Either way, the relative information flow $\tau_{2\to1}$ approaches -1 in
the limit of deterministic system. This is indeed what we expect. So even
for this extreme case, there is no contradiction at all for causal
inference using information flow.

%\section{Discussion}
To summarize, a recent rigorously formulated causality analysis asserts
that, in the linear limit, causation implies correlation, while correlation
does not necessarily mean causation. 
In this short note, a case which seemingly vioates the assertion
is examined. In this case
an event $A$ takes a harmonic form (sine/cosine), and generates through
some process another event $B$ so that $B$ is always out of phase with 
$A$, i.e., lag $A$ by $90^o$. Obviously $A$ causes $B$, but
by computation the correlation between $A$ and $B$ is zero. 
In this study we show that this is an extreme case, with only one point in
the ensemble space and hence the problem becomes singular.
We re-examine the problem by enlarging the ensemble space slightly 
through adding some noise. A stochastic differential equation is then
solved for the corresponding covariances, which allows us to obtaint the
information flows for the perturbed system.
Then as the noisy perturbation goes to zero, 
the normalized information flow rate from $A$ to $B$ is established to 
be 100\%, just as one would have expected. So actually no contradiction
exists.

One thing that merits mentioning is that, here although it seems that $A$
causes $B$, actually here the normalized information flow rate from $B$
to $A$ is also 100\%. That is to say, for such a harmonic system with
circular cause-effect relation, it is actually impossible to differentiate 
causality by simply assessing which takes place first; anyhow, taking lead
by $\pi/2$ is equivalent to lagging by $3\pi/2$.
% Traditionally time-delayed correlation analysis has been widely 
% used in inferring causality, particularly in climate science. 
% However, there has long been argued that correlation is not causation 
% ever since Berkeley (1710). 
The moral is, for a process that is nonsequential 
(e.g., that in the nonsequential stochastic control systems), circular 
cause and consequence coexist, it is essentially impossible to distinguish 
a delay from an advance.

\vskip 0.5cm 

\noindent
{\bf Acknowledgments.}
This study was partially supported by 
the National Science Foundation of China (NSFC) under Grant No.~41975064,
and the 2015 Jiangsu Program for Innovation Research and Entrepreneurship
Groups.

\end{document}